\documentclass[12pt]{article}

\usepackage{newtxtext,newtxmath}

\usepackage{graphicx}

\usepackage[linesnumbered,ruled,vlined]{algorithm2e}

\usepackage{soul}

\usepackage{setspace}
\usepackage[font=small,labelfont=bf]{caption}
\usepackage{afterpage}

\usepackage{float}

\usepackage[letterpaper,margin=1in]{geometry}

\renewenvironment{abstract}
	{\quotation}
	{\endquotation}

\date{}

\makeatletter
\renewcommand{\fnum@figure}{\textbf{Figure \thefigure}}
\renewcommand{\fnum@table}{\textbf{Table \thetable}}
\makeatother

\usepackage{url}

\usepackage{caption}

\usepackage{bm}

\usepackage{listings}

\lstdefinestyle{customverbatim}{
    breaklines=true,                
    breakatwhitespace=true,         
    basicstyle=\scriptsize\ttfamily,           
    columns=fullflexible,           
    frame=single,                   
    postbreak=\mbox{\textcolor{red}{$\hookrightarrow$}\space} 
}
\def\scititle{
    
A multi-agentic framework for real-time, autonomous freeform metasurface design
    
}

\title{\bfseries \boldmath \scititle}

\author{
	Robert~Lupoiu$^{1}$,
	Yixuan~Shao$^{1}$,
    Tianxiang~Dai$^{1}$,
	Chenkai~Mao$^{1}$,
    Kofi~Edée$^{2}$,
    \\Jonathan~A.~Fan$^{1\ast}$\and
	\small$^{1}$Department of Electrical Engineering, Stanford University, Stanford, CA 94305, USA.\and
	\small$^{2}$Université Clermont Auvergne, Institut Pascal, BP 10448, F-63000, Clermont-Ferrand, France\and
	\small$^\ast$Corresponding author. Email: jonfan@stanford.edu\and
}

\begin{document} 

\maketitle

\begin{abstract} \bfseries \boldmath





Innovation in nanophotonics currently relies on human experts who synergize specialized knowledge in photonics and coding with simulation and optimization algorithms, entailing design cycles that are time-consuming, computationally demanding, and frequently suboptimal.  We introduce MetaChat, a multi-agentic design framework that can translate semantically described photonic design goals into high-performance, freeform device layouts in an automated, nearly real-time manner.  Multi-step reasoning is enabled by our Agentic Iterative Monologue (AIM) paradigm, which coherently interfaces agents with code-based tools, other specialized agents, and human designers.  Design acceleration is facilitated by Feature-wise Linear Modulation-conditioned Maxwell surrogate solvers that support the generalized evaluation of metasurface structures.  We use freeform dielectric metasurfaces as a model system and demonstrate with MetaChat the design of multi-objective, multi-wavelength metasurfaces orders of magnitude faster than conventional methods.  These concepts present a scientific computing blueprint for utilizing specialist design agents, surrogate solvers, and human interactions to drive multi-physics innovation and discovery.

\end{abstract}

\subsection*{Introduction}



Optical metasurfaces control light through engineered diffraction, enabling a wide variety of advanced imaging \cite{kim_metasurface-driven_2022, tseng_neural_2021}, display \cite{gopakumar_full-colour_2024, joo_metasurface-driven_2020}, and sensing systems \cite{kim_holographic_2021, guo_compact_2019}. The metasurface design process presents unique opportunities and challenges in scientific computing, as it leverages intricate relationships between subwavelength-scale geometric structure and electromagnetic response  \cite{yu_flat_2014, chen_flat_2020, ni_metasurface_2013, lin_all-optical_2018, mohammadi_estakhri_inverse-designed_2019}. Consequently, significant efforts have been made to navigate this exponentially large design space \cite{jiang_deep_2020, ma_deep_2021, li_empowering_2022, kuang_computational_2020, jiang_global_2019, jiang_simulator-based_2020, kudyshev_machine_2020, arya_end--end_2024}.
Conventionally, photonic design tasks in both academic and industrial settings involve subject experts who apply prior knowledge and experience to propose and simulate candidate device concepts \cite{christiansen_inverse_2021}.  These approaches include heuristic algorithms that leverage physical insights and approximations to efficiently explore the photonic design space \cite{angeris_heuristic_2021, thompson_particle_2021, egorov_genetically_2017, jafar-zanjani_adaptive_2018, yu_genetically_2017}, as well as gradient-based  methods that iteratively refine designs by optimizing a differentiable figure of merit \cite{miller_photonic_2012, sell_large-angle_2017, arya_end--end_2024, pestourie_inverse_2018, piggott_fabrication-constrained_2017, mansouree_large-scale_2021, colburn_metasurface_2018}.  While existing workflows have been effective at advancing the field, they require extensive time and resources to train personnel, ideate, code, and execute fullwave simulation and design tasks.  Furthermore, the need for nanophotonics expertise to perform advanced metasurface design has presented challenges for researchers in adjacent research fields to incorporate detailed metasurface concepts into their workflows.


To this end, deep learning algorithms have been touted as next-generation tools for transforming and streamlining the computer-aided design process in the physical sciences, including in photonics \cite{yeung_deepadjoint_2022, campbell_machine_2023}.  In the realm of mathematical computing, neural networks have been developed to solve partial differential equations using architectures ranging from physics-informed neural networks \cite{raissi_physics-informed_2019, meng_ppinn_2020, kharazmi_hp-vpinns_2021, lu_physics-informed_2021, yuan_-pinn_2022} to Fourier neural operators \cite{li_fourier_2021, kovachki_universal_2021, wen_u-fnoenhanced_2022, li_fourier_2023}, and they have been utilized to effectively perform end-to-end and iterative freeform optimization tasks \cite{chen_high_2022, mao_towards_2024}.  For example, surrogate deep network solvers and optimizers have been demonstrated to accurately evaluate Maxwell's equations and determine structure-function relations with orders of magnitude faster speeds \cite{wiecha_deep_2020, chen_high_2022, zhelyeznyakov_large_2023}.  There have also been significant efforts to develop and integrate large language models (LLMs) within the scientific computing pipeline, to support semantic interpretation of designer needs \cite{ghafarollahi_atomagents_2024, jiang_opticomm-gpt_2024, li_english_2023}, fuse scientific knowledge with the design process \cite{ghafarollahi_atomagents_2024, ghafarollahi_sciagents_2024, kim_nanophotonic_2025}, and assist in detailed coding tasks \cite{li_english_2023}.  To date, LLM frameworks have been used to assist in tasks in a diverse range of physical science topics, such as the design of materials \cite{ghafarollahi_atomagents_2024}, optical fibers \cite{jiang_opticomm-gpt_2024}, and laser cavities \cite{li_english_2023}.  

\begin{figure} 
	\centering
	\includegraphics[width=1.0\textwidth]{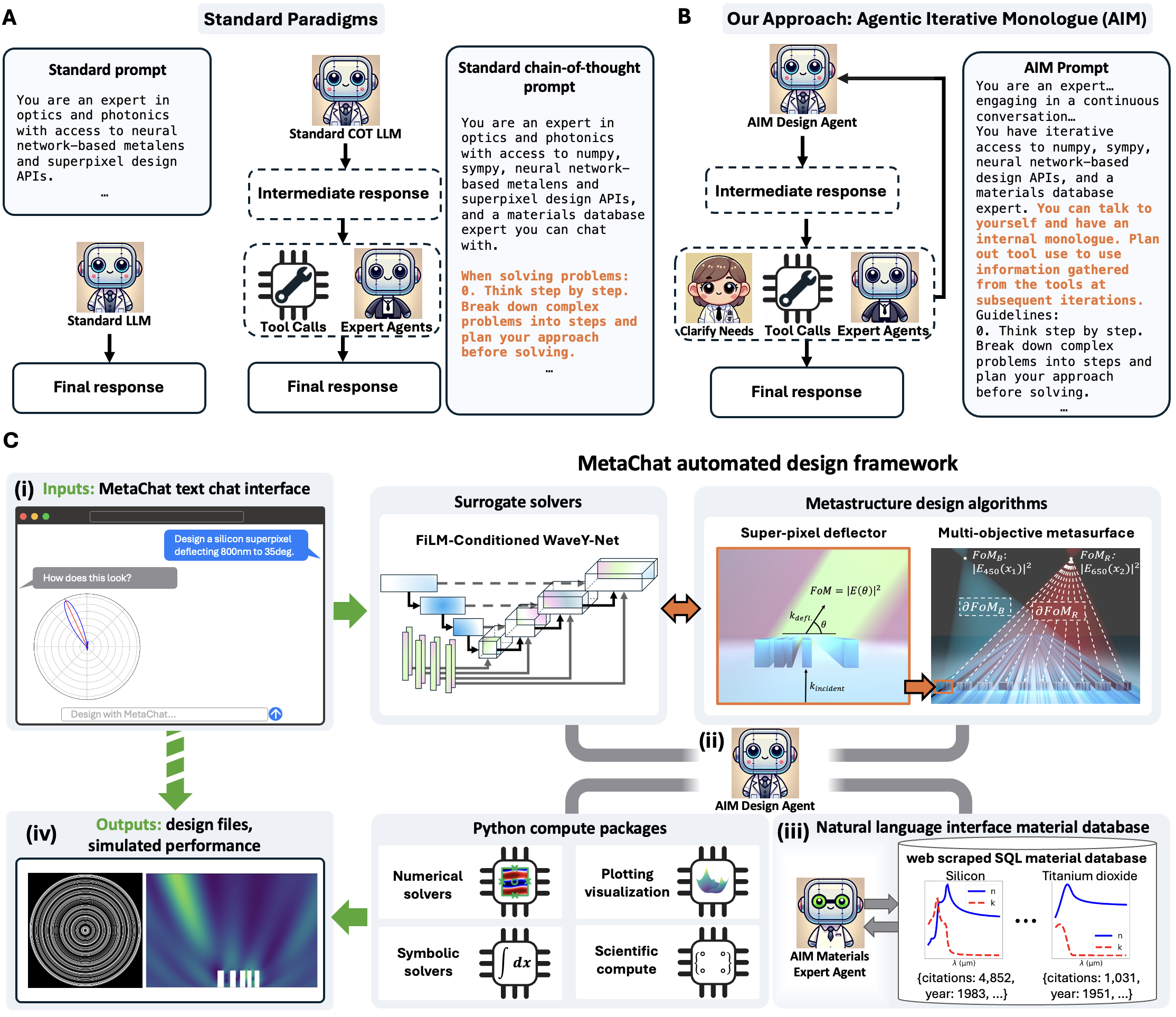} 

	\caption{\textbf{MetaChat agentic framework for autonomous metasurface design.}
        (\textbf{A}) Standard paradigms for creating LLM assistants. (Left) Standard prompting instructs the LLM to assume a certain role for each response, which is generated directly from a given input query. (Right) In chain-of-thought prompting, the LLM is further instructed to explicitly break down the input request into multiple steps before responding, which structures its internal reasoning to help maintain coherence for complex queries. External tool calls can be invoked in a non-iterative, single shot manner.
        (\textbf{B}) The AIM prompting strategy creates agents that refine their intermediate responses through self-dialogue, autonomously leveraging feedback from external tools, agents, and the human designer at multiple stages to optimize the final answer.
        (\textbf{C}) Overview of the MetaChat framework.  AIM Design (ii) and Materials Expert (iii) Agents iteratively interface with high speed metasurface optimization and design algorithms, Python computing packages, and the human user to process semantic user inputs (i) into outputted metasurface design files (iv) in nearly real time.
        }
	\label{fig1} 
\end{figure}

While there has been remarkable progress in augmenting photonics design with the data sciences, it is challenging to extend the functionality of these concepts towards non-trivial use cases.  Deep learning-based mathematical computing algorithms remain highly specialized and cannot account for the diversity of physical parameters required to describe practical photonic systems.  LLM design frameworks for the physical sciences largely serve as planners or wrappers with user-friendly interfaces \cite{jiang_opticomm-gpt_2024, li_english_2023}. These are built around the decomposition of problems into logical steps and the preplanned use of external tools (Fig. \ref{fig1}A) \cite{jiang_opticomm-gpt_2024, ghafarollahi_atomagents_2024}, leading to the rigid and sequential execution of LLM-generated instructions. While many of these tools claim to be agentic, they do not constitute a design framework featuring true \textit{agency}, which embodies intentionality, forethought, self-regulation, and self-reflectiveness \cite{bandura_social_2001}. Thus, they lack the capability of acting on intermediate thoughts or autonomously adapting actions based on feedback from interactions with tools, other agents, or the human user.  

We present for the first time a multi-agentic computer-aided design framework, called MetaChat, which combines true agency with millisecond-speed deep learning surrogate solvers to automate and accelerate photonics design.  We show that MetaChat is capable of performing complex freeform design tasks in nearly real-time, as opposed to the days-to-weeks required by the manual use of conventional computing methods and resources. This immense acceleration in scientific computing capability not only pushes the speed of innovation and design discovery to new levels, but also enables MetaChat to rapidly experiment, autonomously course-correct, and seek feedback from the user on quick time scales that permit practical, interactive human engagement.  These agentic capabilities are enabled by a key concept that we propose, \textit{Agentic Iterative Monologue (AIM)}, a novel agentic system designed to seamlessly automate multiple-agent collaboration, human-designer interaction, and computational tools (Fig. \ref{fig1}B).  We additionally propose a semi-general fullwave surrogate solver, termed Feature-wise Linear Modulation (FiLM) WaveY-Net, which supports conditional fullwave modeling---enabling simulations with variable conditions, including source incident angle, wavelength, material, and device topology---while maintaining high fidelity to the governing physics. We focus MetaChat on the design of two-dimensional dielectric metasurfaces, though the concepts can be readily adapted and generalized to other photonic device domains. With MetaChat, we demonstrate the automated design of multi-objective, multi-wavelength metalenses and deflectors with efficiency levels that match or surpass those previously reported in the literature. 

A more detailed overview of MetaChat is in Fig. \ref{fig1}C.  MetaChat consists of an AIM Design Agent, which performs detailed metasurface design, and an AIM Materials Expert Agent, which is specialized in optical material properties and can perform candidate material identification.
The agents autonomously engage with the human designer through a chat-like interface, with each other, with a variety of external tools, and with themselves through self-thought. In this way, they each continually update their course of action based on intermediate results and feedback. The AIM Design Agent utilizes a suite of application programming interfaces (APIs) to leverage FiLM WaveY-Net and gradient-based freeform optimization algorithms.  The AIM Design Agent also has access to a coding environment equipped with compute packages commonly used by human designers during the design and validation process. These packages include scientific computing tools capable of  rigorous numerical and matrix-based calculations, symbolic solvers that enable  equation manipulation and analytical solving capabilities, conventional numerical Maxwell solvers, and visualization packages that generate graphical outputs for human designer interpretation. The Materials Agent is capable of generating context-dependent responses to questions that typically have multiple viable solutions by using structured query language (SQL) to traverse an information-rich database, overcoming the limitations of traditional lookup table approaches that would suffer from the one-to-many problem. 


\subsection*{Results}

\subsubsection*{Agentic autonomous nanophotonics designers}

The heart of MetaChat is the AIM paradigm for creating LLM-driven agents that excel at scientific reasoning and design (Fig. \ref{fig1}B).  AIM treats each of its outputs as an intermediate thought by default, from which external tool interactions, conversations with other agents, and dialogue with human users are iteratively and autonomously self-invoked using output tags.  AIM builds on \textit{ReAct} and related frameworks \cite{yao_react_2023, fu_preact_2024}, which interleave reasoning within a chain-of-thought trajectory with actions involving external tools, but which are constrained by a rigid response format that diminishes the agent's ability to interact fluidly with human users and other agents.   AIM is particularly well-suited as an agentic design assistant because it is capable of refining and correcting outcomes based on external feedback and is able to converse with expert agents and the human user equally naturally.

AIM derives its agency from a carefully designed prompt, which is presented in the Methods section and detailed in the Supplementary Information. AIM is distinguished from other agentic frameworks \cite{yao_react_2023, fu_preact_2024} by its unconstrained outputs that are each treated as an intermediate \textit{iterative monologue} thought of the agent, with any tool use or chats with other agents being mediated by HTML-like tags. This recursive monologue loop continues until it is broken with a self-outputted response tag, which channels the final response to the user. Within this iterative process, AIM realizes the core factors of true agency \cite{bandura_social_2001}. Like basic assistant LLMs, \textit{intentionality} is achieved by assigning a clearly defined, purpose-driven identity, in this case as an expert assisting users with optics and photonics problems. \textit{Forethought} is realized using established chain-of-thought prompting techniques, which explicitly instruct the LLM to think step-by-step and break down complex problems into a plan before solving. AIM achieves the final two conditions for agency in a flexible manner that is compatible with human designer collaboration by imposing the default state behavior of internal monologue. \textit{Self-regulation} is achieved via the built-in feedback loop that enables the agent to monitor and adjust its processes through tool and optimization output analysis. \textit{Self-reflectiveness} is fulfilled by the internal dialogue through which the agent assesses its own reasoning to refine its decision-making trajectory. AIM's intrinsic self-regulation, facilitated by iterative self-reflection, enables precise agent-driven multi-step scientific computing, circumventing the rigidity associated with executing predetermined plans lacking intermediate result feedback \cite{jiang_opticomm-gpt_2024, ghafarollahi_atomagents_2024} or requiring granular human expert intervention throughout the decision-making process \cite{swanson_virtual_2024}.

AIM-enabled MetaChat is implemented via a full stack application architecture that integrates a local frontend interface for human designers, a local backend for orchestrating business logic, an LLM server, and a GPU-accelerated machine learning (ML) server for hosting ancillary neural networks, such as surrogate solvers. The block diagram in Fig. \ref{fig2}A illustrates the sequentially numbered information flow within MetaChat. The process starts with initiation of the agentic workflow by the user with a query into the local frontend user interface. The query is passed to the \textit{Design Agent} in the local backend and appended to its \textit{context}, which consists of the full conversation history among the user, the Design Agent, and other agents, along with a prompt instructing the LLM about its photonics design assistant persona, available tools, and expected behavior. The Design Agent context is then passed to the LLM server, where next-token prediction iteratively constructs the response string \cite{vaswani_attention_2017, brown_language_2020}. The response includes tagged placeholders, which are programmatically parsed to facilitate Design Agent interactions with other agents, coding tools, and ancillary neural networks, such as surrogate solvers via API calls.  The Design Agent has the ability to iterate on the results with itself, other agents, and tools until it is ready with a response that is passed back to the frontend user interface, where it is processed into a human-readable format.  The response also includes hyperlinks to the backend database where any solutions are stored.

\begin{figure}[p] 
	\centering
	\includegraphics[width=1.0\textwidth]{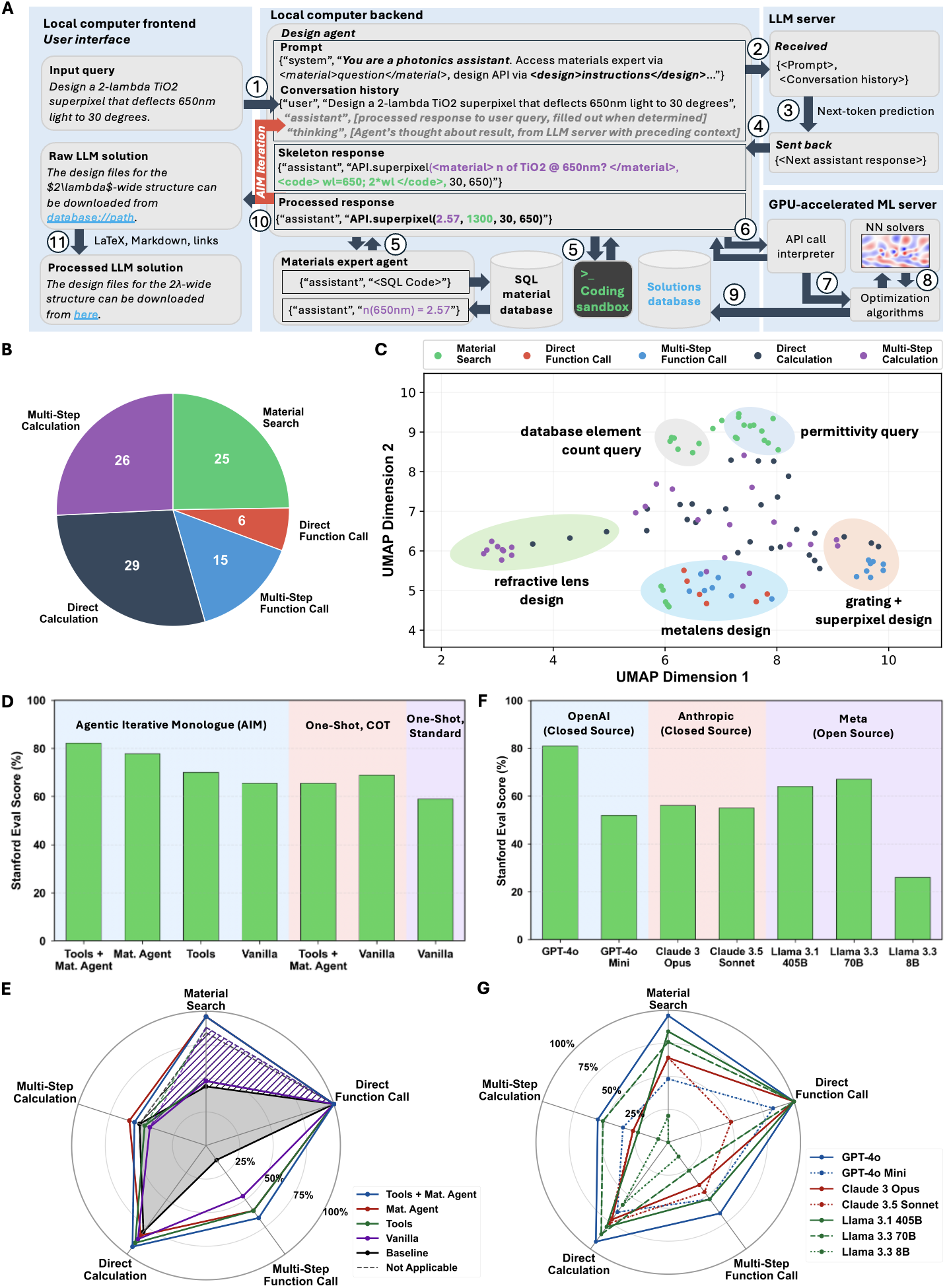} 
\end{figure}

\afterpage{
\clearpage
\begin{center}
\begin{minipage}{\textwidth}
\noindent
\captionof{figure}{\textbf{Photonics design agent evaluation on a graduate-level nanphotonics knowledge and optimization benchmark.}
(\textbf{A}) Block diagram detailing the sequentially numbered flow of information in a Metachat query. User queries are submitted through the local frontend and processed by the backend with the assistance of local databases and coding tools. The backend communicates with externally hosted LLM and GPU-accelerated servers for natural language processing and near real-time numerical optimization, respectively.
(\textbf{B}) Problem category distribution in the Stanford graduate student-generated benchmark.
(\textbf{C}) 2D UMAP projection of all 101 problem embeddings. Each plotted problem is color-coded based on its benchmark category. 
(\textbf{D}) AIM and LLM assistant ablation study using the Stanford  nanophotonics benchmark. Green bars indicate the percentage of correct answers. 
(\textbf{E}) AIM ablation study across the five problem categories of the Stanford nanophotonics benchmark. 
AIM, equipped with both tools and access to the Materials Expert Agent, achieves the highest scores consistently across all five categories, with the most significant improvement observed in the Multi-Step Function Call category. 
(\textbf{F}) AIM LLM driver performance evaluation using the Stanford  nanophotonics benchmark. Bars indicate the percentage of correct answers, with the best-performing LLM being the closed-source GPT-4o (81\%). 
(\textbf{G}) AIM LLM driver performance evaluation across the five problem categories of the Stanford nanophotonics benchmark. GPT-4o is the best-performing model, and it is the only one to achieve consistently high scores across all five categories.
}
\label{fig2}
\end{minipage}
\end{center}
}

To evaluate the performance of our AIM designer, we establish and utilize the Stanford nanophotonics benchmark. This benchmark was developed due to the lack of standardized evaluation metrics for photonics design, and it comprises 101 diverse problems in photonics, electromagnetics, and optical engineering distributed in five categories: Direct Calculation, Multi-Step Calculation, Material Search, Direct Function Call, and Multi-Step Function Call (Fig. \ref{fig2}B).  Each problem is carefully designed to test framework capabilities in a standalone, autonomous environment without human designer input. Database-specific problems, such as element counting or citation reporting, are an important part of the benchmark because they evaluate the multi-agentic capability of MetaChat's Design Agent to delegate problems correctly to the Material Expert Agent, and the latter's proficiency in generating accurate SQL queries.  The distribution of the problems is visualized in Fig. \ref{fig2}C by plotting the first two Uniform Manifold Approximation and Projection (UMAP) dimensions of the question string embeddings \cite{mcinnes_umap_2020, brown_language_2020}. Several clusters of common problems present in the dataset emerge that are commonly encountered by nanophotonics engineers, including metalens design and material permittivity queries. The center of Fig. \ref{fig2}C comprises 37 points that represent general photonics engineering problems that cannot be grouped with a particular commonly encountered class. The structured question-answer format of the benchmark enables the comprehensive and quantitative assessment of the performance of agentic frameworks and their underlying LLMs.


We first use the the evaluation benchmark to quantify the performance of AIM with respect to alternative assistant LLM approaches, and to assess the utility of MetaChat's supporting tooling via ablation studies (Fig. \ref{fig2}, D and E).  GPT-4o is used as a common LLM backbone.  In this assessment, database element count query questions are withheld due to their specificity to the Materials Agent Database, but we note that the AIM Materials Expert Agent is able to address these problems with 100\% accuracy.  The baseline score of a vanilla LLM assistant that outputs an answer directly (Fig. \ref{fig1}A, left), is 65\%.  Through the vanilla chain-of-thought (COT) prompting strategy \cite{wei_chain_2022}, without access to any external tools or the Materials Agent, the score increases to 72\%. Unexpectedly, this score drops to 59\% after tools and Materials Agent access are added to the COT assistant (Fig. \ref{fig1}A, right). This drop in performance is attributed to the requirement for a more complex prompt containing tools and Material Agent use instructions, which the one-shot LLM cannot leverage to its advantage due to diminished instruction-following performance with longer prompt size and errors arising from tool interactions on queries the LLM could otherwise answer directly.

Our AIM designer (Fig. \ref{fig1}B), on the other hand, is able to take advantage of external tools and agents to more accurately solve problems.
The base score of AIM without any external helper tools or Materials Agent access is 71\% and comparable to that of vanilla COT, despite the fact that our AIM designer utilizes a significantly longer and more complex prompt than that used for the vanilla COT assistant. Its performance further improves after adding either tools (75\%) or the Materials agent (78\%), and with both combined external resources, the AIM designer achieves a maximum score of 81\%. With access to both tools and the Materials Agent, the AIM designer consistently achieves high scores across all five categories of the evaluation benchmark (Fig. \ref{fig2}E). AIM outperforms ablated versions and baseline measures most in the multi-step function call category of benchmark problems (Fig. \ref{fig2}E), indicating a higher propensity to maintaining long-term coherence in the face of multi-step calculations and function calls. In particular, the agentic approach succeeds in cases where the pre-planned, one-shot COT assistant encounters tool-calling errors during multi-step function and API calls, due to AIM's ability to interpret error messages and feedback to dynamically adjust its strategy, and to iteratively retry interactions with external tools. These results underscore the importance of the self-regulation property of a truly agentic system, which is a capability not present in rigid pre-planning or wrapper-type approaches to interfacing with design tools.


To evaluate the impact of LLM type on AIM designer performance, we perform our evaluation benchmark on seven different state-of-the-art LLM models.  This assessment is important to perform for our application, as most LLM model benchmarks focus on single-shot, direct responses that are reflective of more typical chatbot-type tasks that are grounded by a human at each step \cite{chiang_chatbot_2024, zheng_lmsys-chat-1m_2024}.  In contrast, our AIM designer has particular demands in instruction-following, function calling, context size, and long-term coherence for multi-step scientific computing task automation. The results are summarized in Fig. \ref{fig2}, F and G.  We determine that OpenAI's closed-source GPT-4o outperforms all tested models. Meta's open-source Llama 3.3 70B model achieves a score of 67\%, which comes the closest to the state-of-the-art score achieved by GPT-4o. These results indicate that model size is not predictive of agentic driver performance. We further find that performance does not track with commonly well-regarded LLM benchmarks.  For example, Anthropic's closed-source Claude 3.5 Sonnet, which is widely regarded as a top-performing LLM model \cite{chiang_chatbot_2024}, performs significantly worse than GPT-4o, Llama 3.1 405B, and Llama 3.3 70B.  In addition, Anthropic's models consistently underperform both GPT-4o and Llama 3.3 70B in more complex multi-step function calls,  as well as in multi-step and direct calculations (Fig. \ref{fig2}G). These observations indicate the importance for our AIM agents to interface with LLM models that support long-term coherence and excellent function-calling capability in order to agentically achieve correct answers. We hypothesize that the better-performing LLMs in Fig. \ref{fig2}F have a relative over-representation of function call and numerical calculation examples in their respective training processes.

\subsubsection*{Ultrafast electromagnetic simulations with FiLM WaveY-Net}

The high-speed surrogate fullwave solver is the mathematical computing engine driving MetaChat.  New deep learning architectures are required, as existing direct deep learning surrogate Maxwell solvers are too specialized and unable to support generalized capabilities to the degree required for practical MetaChat functionality.  We propose FiLM WaveY-Net, a semi-general fullwave solver of metasurface \textit{superpixels} that supports the ultra-high speed simulation of freeform aperiodic dielectric metasurfaces (Fig. \ref{fig3}A).  Superpixels are defined to be wavelength-scale metasurface scatterer sections (Fig. \ref{fig3}B) that can be stitched together to produce a large area, functional freeform metasurface \cite{phan_high-efficiency_2019, pestourie_inverse_2018}.  Compared to strategies based on the stitching of meta-atoms, our superpixel strategy accounts for near-field coupling between neighboring nanostructures and their arbitrary positioning, making it ideal for freeform metasurface design tasks.  To train FiLM WaveY-Net, $270,000$ examples of ground truth training data are produced using a finite-difference frequency domain (FDFD) simulator \cite{hughes_forward-mode_2019}. The superpixels comprise dielectric ridges placed over a glass substrate, and PML boundary conditions are utilized at all sides of the simulation domain (Fig. \ref{fig3}B). The superpixel parameters are uniformly sampled from the ranges listed in Fig. \ref{fig3}C, with half of the dataset positioning the illumination source below the superpixel structure within the glass substrate, and the other half above the structure, in air.

\begin{figure} 
	\centering
	\includegraphics[width=1.0\textwidth]{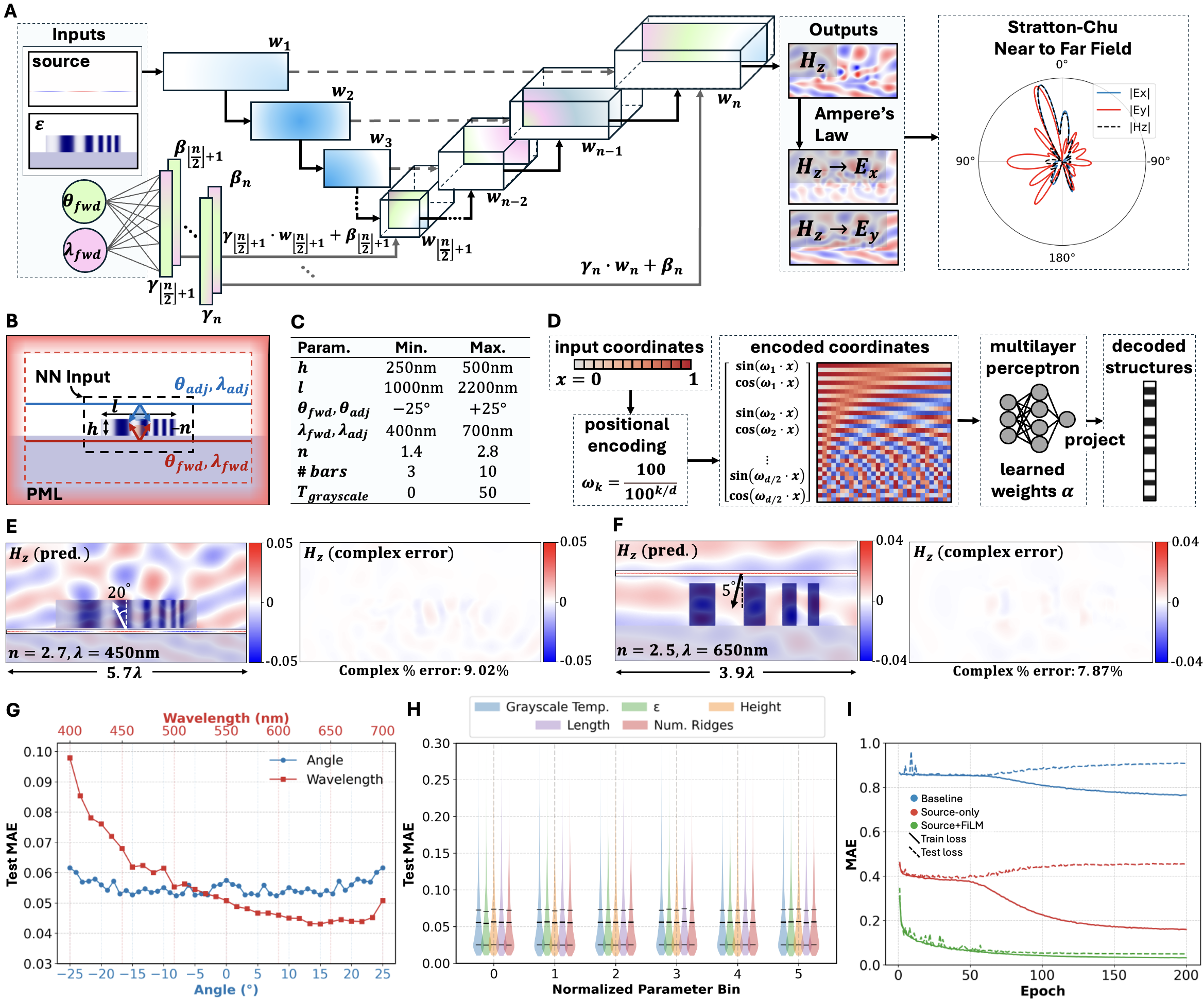} 

	\caption{\textbf{FiLM WaveY-Net fullwave surrogate solver, with support for variable sources and structures.}
    (\textbf{A}) FiLM WaveY-Net is composed of a UNet backbone that uses learned affine transformations on the decoding blocks to assist with conditioning on variable source angles and wavelengths. The UNet input channels include the dielectric distribution and the real and imaginary components of the source. The network output is the magnetic near-field profile, from which electric near-field profiles and far-field scattering profiles via the Stratton-Chu formalism are computed. 
    (\textbf{B}) The input to FiLM WaveY-Net in the context of the simulation domain. The red dashed line indicates the location of the PML boundary condition. The black dashed line indicates the center section of the domain that is inputted into the CNN. The solid red line indicates the location of forward sources, and the solid blue line indicates the location of backward sources. 
    (\textbf{C}) The ranges of the variable quantities in the FiLM-Conditioned WaveY-Net training set. 
    (\textbf{D}) Overview of the differentiable neuroparameterization framework for defining superpixel geometry. The input spatial coordinates $x$ are mapped into a higher-dimensional space using rotary positional encoding. The encoded coordinates are then passed through a multilayer perceptron (MLP), which learns an implicit signed distance function (SDF) representation of the design. 
    (\textbf{E,F}) Two examples of FiLM WaveY-Net surrogate simulation results (left) and complex percentage error between the predicted and ground truth magnetic field (right). 
    }
	\label{fig3} 
\end{figure}

\begin{figure} \ContinuedFloat
    \caption{(continued) 
     (\textbf{G}) Test loss as a function of the variable source parameters: incidence angle (blue) and wavelength (red). The test loss remains stable across the entire angle range but decreases with increasing wavelength due to the reduced complexity of the field profile at lower frequencies.
     (\textbf{H}) Violin plot distributions of test loss across the full range of structure-related input parameters: grayscale temperature (blue), maximum permittivity (green), height (orange), length (purple), and number of ridges (red). Each parameter range is divided into six bins for direct comparison. The bottom quartile, median, and top quartile are indicated with black horizontal lines. The test loss is consistent across all binned parameter ranges, with at least 90\% of the test samples exhibiting an MAE below 0.10 in each distribution. 
     (\textbf{I}) Loss curves for the ablation study of the FiLM-Conditioned WaveY-Net. The ``Baseline" model inputs the structure only; the ``Source-only" model  inputs the source along with the structure; the ``Source+FiLM" model uses the FiLM conditioning technique in conjunction with the source inputted via separate CNN channels alongside the structure to achieve a low loss that generalizes to the test set. 
        }
\end{figure}

The FiLM WaveY-Net architecture builds on previous surrogate frequency domain fullwave solver architectures \cite{wiecha_deep_2020, chen_high_2022, lim_maxwellnet_2022, zhelyeznyakov_large_2023}, where it was shown that convolutional neural networks could be trained to accurately simulate dielectric metasurface structures, but which were limited to highly specific input source conditions.
With FiLM conditioning \cite{perez_film_2017}, the source wavelength and incident angle are inputted as explicit scalar inputs and are converted into feature-wise affine modulation parameters via learned linear transformations.  These parameters modulate intermediate feature maps at each decoding block in the base UNet architecture. Spatial information about the source position is provided through two additional input channels corresponding to the real and imaginary components of the source, alongside the original dielectric structure channel, to the convolutional neural network.  
A key feature of FiLM conditioning that enables generalized solving capabilities is it enables the central convolutional block's encoded dielectric structure representation to be adaptively decoded into a source-specific field response. Functionally, FiLM conditioning facilitates precise source dependency by effectively projecting generalized decoding parameters onto distinct hyperdimensional subspaces that are unique for each illuminating source. 
We use FiLM conditioning to extend our surrogate solving capabilities to the accurate evaluation of a wide range of device heights, binary and grayscale dielectric material configurations, device topologies, illumination wavelengths, and illumination angles (Fig. \ref{fig3}C).  

Importantly, FiLM WaveY-Net can be configured within a fully differentiable end-to-end pipeline for the simulation of freeform superpixels with customized far-field profiles, enabling the use of autodifferentiation to perform freeform gradient-based superpixel design.  Superpixel layouts are parameterized using a novel neuroparameterization approach illustrated in Fig. \ref{fig3}D, which encodes the geometry in a high-dimensional neural signed distance function (SDF) representation. This approach enforces geometric constraints, such as minimum feature size, without excessively limiting the optimization space. To enforce geometric constraints, geometric loss terms are integrated into the optimization process and guide adjustments to the superpixel geometry through gradient backpropagation, thereby leveraging the differentiable nature of the neural SDF. Implementation details are provided in the Methods section and will be further analyzed in future studies.
Evaluation of the superpixel electromagnetic response is subsequently performed with FiLM WaveY-Net, which outputs the magnetic near-field response, and the electric near fields are computed using a discretized form of Ampere's law.  The complex scattering far-field response based on these near fields is evaluated using the Stratton-Chu formalism (Fig. \ref{fig3}A, right). The neural networks, Ampere's law, and the Stratton-Chu calculations are all fully differentiable, enabling the use of gradient-based optimization via backpropagation to iteratively modify the superpixel geometric layout in a manner that pushes its complex far-field response towards a desired pattern.




%
%

An evaluation of FiLM WaveY-Net showcases its generalizability and an accuracy level that is sufficiently high for inferred fields to be directly used in gradient-based optimization algorithms.  An inspection of two randomly sampled structures and illumination conditions is shown in Fig. \ref{fig3}E and \ref{fig3}F, illustrating the quality and low error level of representative outputted fields.  A more systematic evaluation of FiLM WaveY-Net using a test set of $30,000$ examples provides quantitative performance metrics.  The angle-dependent test performance exhibits a steady normalized mean absolute error (MAE) averaging $0.055$, with a slight increase to $0.062$ at the extremity of the training set (Fig. \ref{fig3}G). This level of consistency indicates that varying the source angle does not pose variability in training difficulty for FiLM WaveY-Net. The representation of wavelength data in the training set is dynamically tuned based on the wavelength-dependent test performance during training to account for variability in learning difficulty, which is due to the increase in spatial information from higher source frequency illumination. Nonetheless, the final test MAE exhibits an inverse relationship with respect to wavelength, ranging from $0.098$ to $0.043$ across the $400\, \text{nm}$ to $700\, \text{nm}$ range (Fig. \ref{fig3}G). The test performance is consistent across the variable dielectric structure parameters, with each parameter exhibiting the same mean of $0.06$ normalized MAE across its entire parameter range, and with $90\%$ of samples exhibiting a loss under $0.10$ (Fig. \ref{fig3}H). 

Ablation studies highlight the essential role that our FiLM conditioning mechanism plays in ensuring robust network generalization to diverse source conditions (Fig. \ref{fig3}I). In the baseline evaluation without any explicit conditional input, the vanilla WaveY-Net architecture \cite{chen_high_2022} is incapable of learning this ill-posed, one-to-many training task (Fig. \ref{fig3}I, blue traces). We also perform a benchmark evaluation in which source information is explicitly included in the channel input, making the training problem well-posed.  Nonetheless, this straightforward conditioning method is insufficient and the resulting trained networks exhibit significant training set overfitting (Fig. \ref{fig3}I, red traces). Related conditioning strategies in which more explicit, learned feature maps are injected into the network also fail to generalize beyond the training set, as detailed in the Supplementary Information. Indeed, FiLM WaveY-Net is qualitatively better and accurately maps conditional fields to corresponding input structures and sources without overfitting (Fig. \ref{fig3}I, green traces). 

\subsubsection*{Autonomous design of multi-wavelength, multi-objective metasurfaces}

The programming interface that links the MetaChat AIM agents with FiLM WaveY-Net are APIs that are tailored to translate a desired device function to a set of optimization objectives and fullwave solver calls. For this study's demonstrations, we limit the APIs to multi-functional metasurfaces serving as deflectors and lenses, though the design and API frameworks can readily extend to more arbitrary wavefront responses.  Deflectors are specified to produce transmitted linear phase responses while lenses are specified to produce transmitted parabolic phase profiles. While these optical functions are relatively basic, they are good model systems for our study, as it is highly non-trivial to effectively design efficient multi-functional devices with these capabilities.
The API-enabled design strategy is shown in Fig. \ref{fig4}A.  First, global metasurface phase responses are specified and quantified based on the desired device function.  In the case of multi-functional devices, sets of phase responses are specified for differing incident angles and wavelengths.  Second, these spatial phase profiles are mapped to desired superpixel far-field scattering profiles, which stitch together to specify the full metasurface response.  The corresponding global Figure of Merit (FoM) to be minimized, which produces the desired phase responses, is the negative sum of the complex electric field amplitudes for each superpixel at points in the far field where intensity is maximal. Our approach streamlines and generalizes multi-objective metasurface design, as all tasks map onto the same superpixel scattering profile optimization approach.

\begin{figure} 
	\centering
	\includegraphics[width=1.0\textwidth]{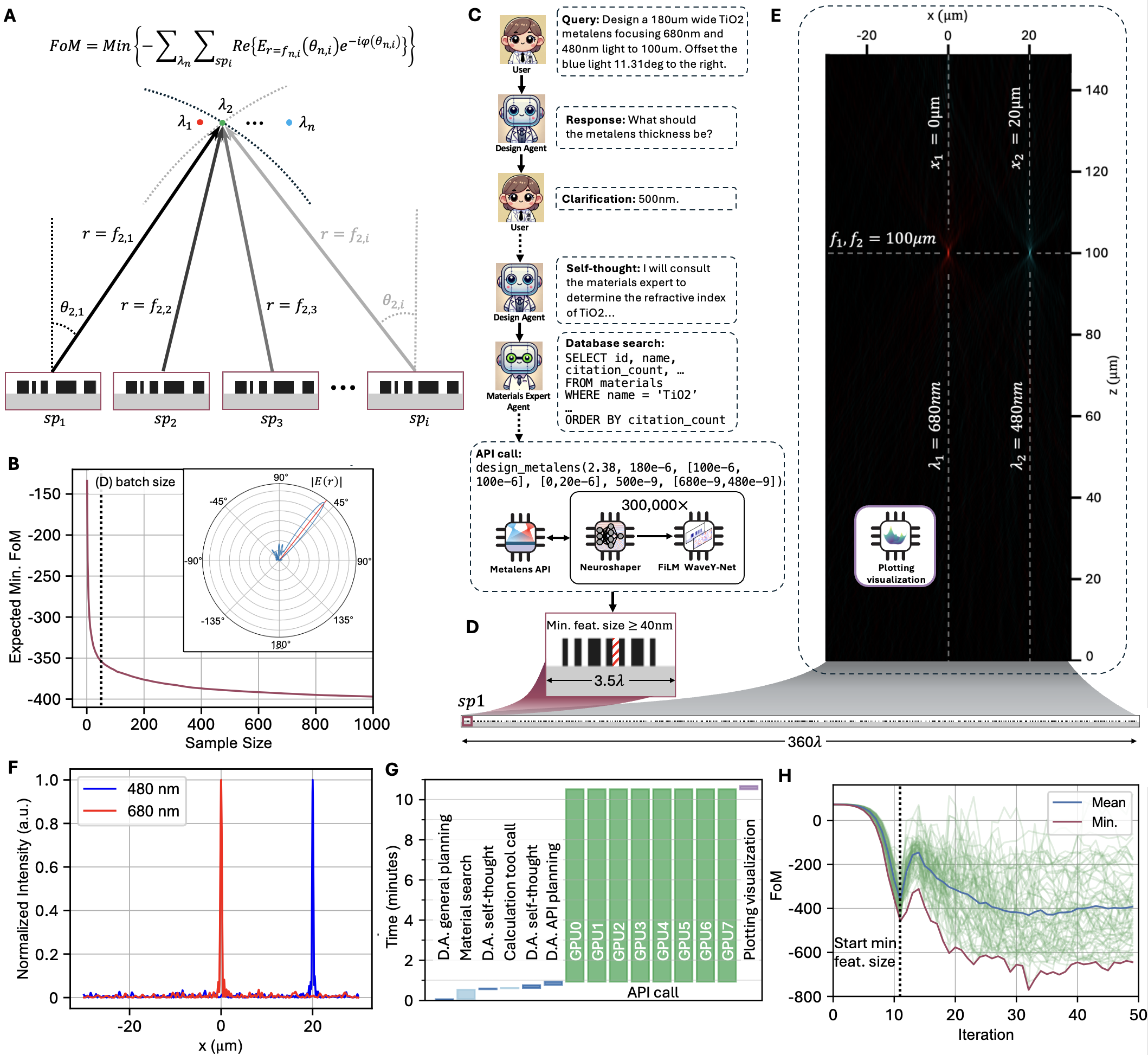} 

	\caption{\textbf{Automated multi-objective metalens design via MetaChat.}
    (\textbf{A}) Design strategy of metasurfaces using superpixel arrays.  
 The FoM captures the electric field magnitudes at the desired focal point locations, which is maximized by the constructive interference of complex scattering fields from all superpixels.
 (\textbf{B}) Expected minimum FoM with respect to sample size based on Monte Carlo sampling from 1,000 optimization runs. We observe a logarithmic improvement in FoM with increased superpixel batch size, and that batches of 60 structures balances performance with compute time. The inset depicts the best superpixel far-field scattering profile (target amplitude: red line) from a batch size of 60, showing minimal side lobes.
 (\textbf{C}) Salient speech excerpts from the design conversation between the User, Design Agent, and Materials Expert Agent for the design of a dual-wavelength, multi-functional metasurface. The Design Agent asks for clarification before engaging in self-thought and autonomous conversation with the Materials Expert Agent to gather the information needed to design a custom dual-objective metalens spanning 360 wavelengths. 
 (\textbf{D}) The final metalens design, which comprises 100 superpixels that each are $\sim\! 3.5$ wavelengths wide. 
        }
	\label{fig4} 
\end{figure}

\begin{figure} \ContinuedFloat
    \caption{(continued)
    (\textbf{E}) The spectral color mapped power profile of the optimized metalens, plotted using the MetaChat visualization tool. The horizontal dashed line indicates the focal plane and the vertical dashed lines indicate the red and blue focal spots.
    (\textbf{F}) Normalized intensity in the metalens focal plane as a function of lateral position, featuring a $480\, \text{nm}$ peak at $20\,\upmu\text{m}$ and a $680\, \text{nm}$ peak at $0\,\upmu \text{m}$.
    (\textbf{G}) Execution time profiling of the design process. Design Agent iterations are colored in dark blue and tool calls are in light blue. The API call compute is in green and plotting visualization is in purple. This entire MetaChat design process takes 11 minutes, with the 300,000 FiLM WaveY-Net simulations accounting for 10 minutes.
    (\textbf{H}) The 60 parallel optimization trajectories for one of the superpixels comprising the metalens. The start of minimum feature size enforcement is indicated via the vertical dashed line and is accompanied by a temporary bounce in the FoM. The trajectory that leads to the best-performing structure that is ultimately included in the metalens design is the red line. The mean FoM is the blue line.
        }
\end{figure}

Importantly, our global FoM can be readily decomposed into independent superpixel FoMs that each specify an individual scattering profile design task, thereby enabling the independent and parallel design of each superpixel.  Details pertaining to individual gradient-based superpixel optimization are in the Methods section.  Briefly, superpixels are initialized using our neuroparameterization framework and are specified as a random, high-temperature grayscale dielectric distribution.  This distribution is evolved by iteratively evaluating the complex superpixel scattering profile using FiLM WaveY-Net and Stratton-Chu calculations, then using autodifferentiation and backpropagation to calculate gradients to the dielectric distribution in a manner that pushes the far-field response towards the desired profile. We use Adam gradient descent \cite{kingma_adam_2015}, and the temperature of the dielectric distribution is programmatically lowered over the course of optimization so that the final superpixels have discrete dielectric values.  

Our gradient-based optimization method is ultimately a local optimizer that is sensitive to its initialization, thus improved results can be obtained by optimizing batches of superpixels with differing initializations and selecting the best structure.  Such a strategy aligns with the strength of FiLM WaveY-Net, which is high speed and leverages GPU-supported batch parallelization.  A Monte Carlo sampling analysis of superpixel FoM as a function of batch size for a typical design task (Fig. \ref{fig4}B), where the best final FoM achieved within a batch is plotted, shows the inherent tradeoff between device performance and computing resources.  This tradeoff can be explicitly accounted for by the Design Agent depending on human designer needs.

To illustrate how the AIM Design Agent interfaces with our general API design framework, we task MetaChat to design a dual-wavelength metalens (Fig. \ref{fig4}C).  We input the plain language design query: ``Design a $180\, \upmu \text{m}$ wide TiO2 metalens focusing $680\, \text{nm}$ and $480\, \text{nm}$ light to 100um. Offset the blue light 11.31deg to the right" (Fig. \ref{fig4}C, Top).  The Design Agent queries the user for missing thickness constraint information, followed by a string of self-thoughts that lead to a Materials Expert Agent query. This chain of actions culminates in an API call that sets off 300,000 GPU-parallelized FiLM WaveY-Net simulations for the gradient-based optimization of a set of 100 superpixels (Fig. \ref{fig4}C, Bottom).  Each of the superpixels spans $1.8\, \upmu \text{m}$ in length and is constrained to support a default $40\, \text{nm}$ minimum feature size, enforced by our neuroparameterization method (Fig. \ref{fig4}D, Top). Once the superpixel optimization process is completed, the superpixels are stitched together to form a fully aperiodic metalens (Fig. \ref{fig4}D, Bottom). With MetaChat's visualization tools, the metasurface far field is plotted using the angular spectrum method (Fig. \ref{fig4}E). As designed, the focal plane is situated at $100\, \upmu \text{m}$, with the $680\, \text{nm}$ focal spot centered and the $480\, \text{nm}$ focal spot offset by $20\, \upmu \text{m}$. The intensity along the focal plane is plotted in Fig. \ref{fig4}F and shows that the $680\, \text{nm}$ and $480\, \text{nm}$ peak lobes contain $57.1\%$ and $49.7\%$ of their respective total far-field energy scattering profile. These results compare favorably with the performance of similar devices from previously reported related demonstrations \cite{pestourie_inverse_2018, chen_gan_2017}.

The timing analysis for the entire agentic design process of the metalens is presented in Fig. \ref{fig4}G. The agentic interactions, including self-thinking and Materials Expert conversation, are complete within the first minute. With the computing resources allocated for this design problem, 8 GPUs are utilized in parallel to complete the required 300,000 simulations in 10 minutes, after which plotting the final result is complete within 30 seconds. A similar metalens would take a knowledgeable practitioner 5.03 days in compute time alone using a classical FDFD simulation approach on an 80-thread server \cite{hughes_forward-mode_2019}. A plot of the FoM of 60 optimizations as a function of iteration number for a representative superpixel type is shown in Fig. \ref{fig4}H and shows the wide variance in superpixel performance given random device initialization and the need to parallelize large optimization batches to identify suitably high performing device structures.

In a second demonstration, we use MetaChat to optimize a multi-objective, large numerical aperture metalens operating at red, green, and blue (RGB) wavelengths with the query in Fig. \ref{fig5}A. The $540\,\text{nm}$ green focal point is instructed to be centered, and the $670\,\text{nm}$ red and $450\,\text{nm}$ blue focal points are instructed to be offset by $30\,\upmu\text{m}$ to the left and right, respectively. The focal plane is specified to be $100\, \upmu\text{m}$ above the device for each spot. The metalens is specified to be $200\,\upmu\text{m}$ wide and is decomposed into 111 superpixels. The far field of the optimized metalens is plotted in Fig. \ref{fig5}B, illustrating the design's success in shaping the wavefront of the three wavelengths into their intended focal points. The field intensity along the focal plane is plotted in Fig. \ref{fig5}C, indicating a state-of-the-art level of signal to noise ratio at the correctly positioned focal points \cite{pestourie_inverse_2018, chen_gan_2017}. 

\begin{figure} 
	\centering
	\includegraphics[width=1.0\textwidth]{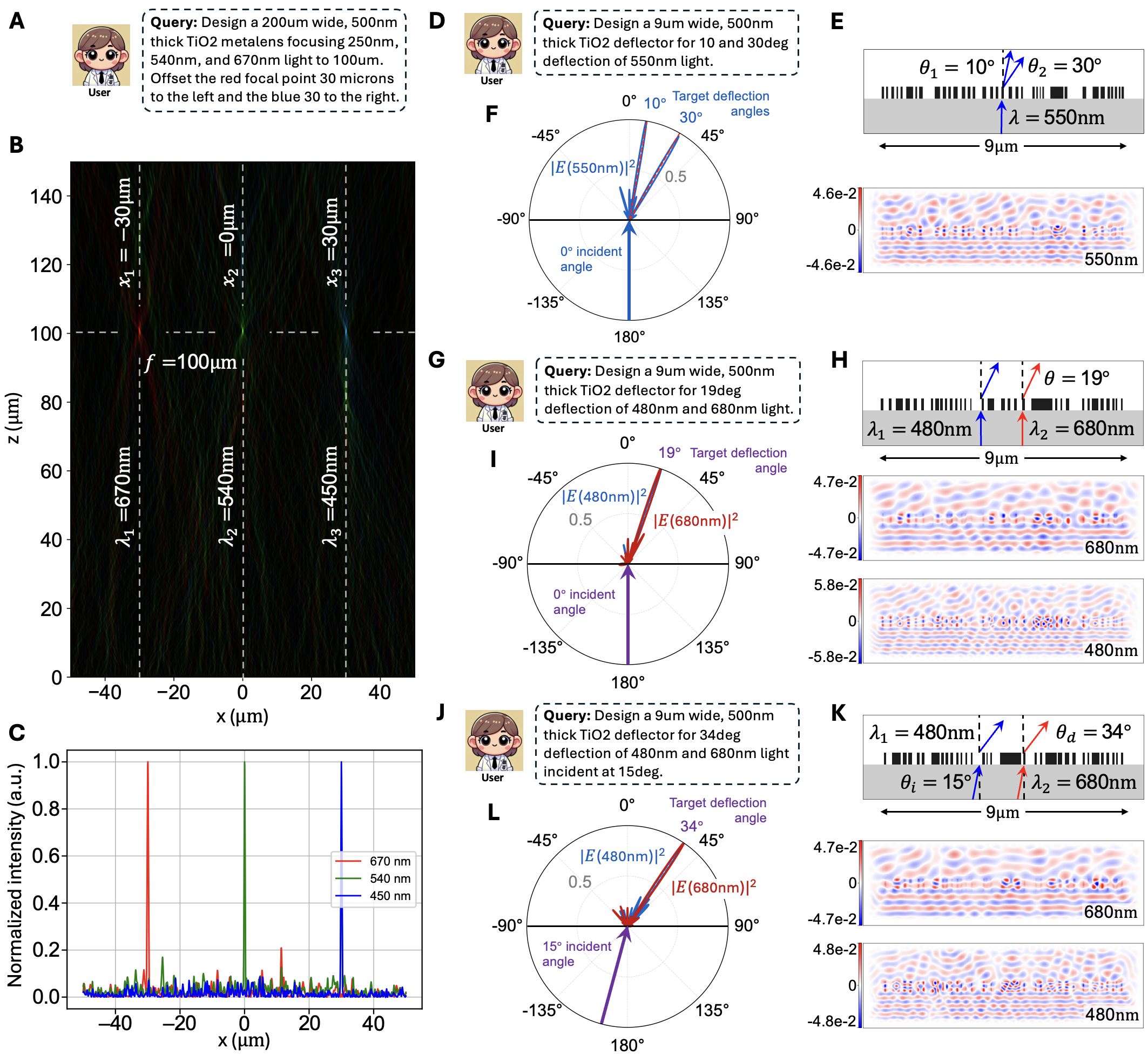} 

	\caption{\textbf{Autonomous optimization of multi-wavelength, multi-objective metalenses and deflectors.}
        (\textbf{A}) User design query for the autonomous MetaChat design of an RGB metalens with a $30\,\upmu\text{m}$ lateral shift between the three focal points. 
        (\textbf{B}) Spectrally color-mapped far-field power profile of the optimized metalens, plotted using MetaChat visualization. The dashed horizontal line indicates a common focal length of $100\,\upmu\text{m}$, and the dashed vertical lines indicate lateral positions of $-30\,\upmu\text{m}$, $0\,\upmu\text{m}$, and $30\,\upmu\text{m}$ relative to the center for the red, green, and blue focal points, respectively.
        (\textbf{C}) Normalized intensity in the metalens focal plane as a function of lateral position for the device in (B).
        (\textbf{D}) User design query for the autonomous MetaChat design of a dual-peak, single-wavelength beam deflector with normal incidence.
         (\textbf{E}) (Top) $550\,\text{nm}$ light is normally incident on the five stitched binarized superpixels optimized via MetaChat to simultaneously deflect the light to $10^\circ$ and $30^\circ$. (Bottom) FDFD-calculated near-field response.
        (\textbf{F}) Far-field response of the five-superpixel deflector. The only two prominent scattering peaks are centered on the designed angles of $10^\circ$ and $30^\circ$.
        (\textbf{G}) User design query for the autonomous MetaChat design of a dual-wavelength, single-direction beam deflector with normal incidence.
        }
	\label{fig5} 
\end{figure}

\begin{figure} \ContinuedFloat
    \caption{(continued) 
        (\textbf{H}) (Top) $480\,\text{nm}$ and $680\,\text{nm}$ light is normally incident on the five stitched binarized superpixels optimized via MetaChat to simultaneously deflect both beams to $19^\circ$. (Middle, Bottom) FDFD-calculated near-field responses of the optimized structure for both wavelengths.
        (\textbf{I}) Far-field response of the five-superpixel deflector. The only prominent peaks of the normally incident $680\,\text{nm}$ and $480\,\text{nm}$ light beams in the far field are both centered on the single design angle of $19^\circ$.
        (\textbf{J}) User design query for the autonomous MetaChat design of a dual-wavelength, single-direction beam deflector with oblique incidence.
        (\textbf{K}) (Top) $480\,\text{nm}$ and $680\,\text{nm}$ light is obliquely incident at $15^\circ$ on the five stitched binarized superpixels optimized via MetaChat to simultaneously deflect both beams to $34^\circ$. (Middle, Bottom) FDFD-calculated near-field responses of the optimized structure for both wavelengths.
        (\textbf{L}) Far-field response of the five-superpixel deflector. The only prominent peaks of the obliquely incident $680\,\text{nm}$ and $480\,\text{nm}$ light beams in the far field are both centered on the single design angle of $34^\circ$.
        }
\end{figure}

Finally, we demonstrate that MetaChat can be used to design highly directional multi-objective, multi-wavelength deflectors. For example, a beam splitter with arbitrarily engineered directions (Fig. \ref{fig5}E, top) can be designed with the user query from Fig. \ref{fig5}D. The resulting field is simulated using FDFD (Fig. \ref{fig5}E, Bottom) and the far-field intensity is plotted using the Stratton-Chu near-to-far field transformation (Fig. \ref{fig5}F), showing that the normally incident light is simultaneously deflected to $10^\circ$ and $30^\circ$. MetaChat can readily generalize these design tasks for sources that are normally incident (Fig. \ref{fig5}, G and H) and obliquely incident (Fig. \ref{fig5}, J and K) onto the metasurface. Both of these examples exhibit the desired far-field  behavior (Fig. \ref{fig5}, I and L) with favorable far-field efficiency levels compared to previously reported metasurface designs sharing similar objectives \cite{choi_multiwavelength_2024}. 

\subsection*{Discussion}



Nanophotonic devices, such as metasurfaces, have faced longstanding barriers to widespread adoption, limiting accessibility primarily to domain experts due to the specialized knowledge, computational demands, and rigid optimization schemes that are traditionally required \cite{christiansen_inverse_2021}. In this work, we introduce MetaChat, a multi-agentic design platform that lays the foundation for agentic automated photonic computer-aided design. MetaChat leverages AIM, a novel paradigm for LLM-driven agency, to automate this computationally intensive process that is traditionally reliant on domain-specific experts. On a graduate-level agentic design benchmark for photonics, AIM demonstrates long-range coherence across multi-step design problems that includes diverse scientific computing tool use and fluid human designer interactions. We show that the AIM Design Agent within the MetaChat framework is capable of achieving best-in-class designs for multi-objective, multi-wavelength metalenses and metasurface deflectors \cite{choi_multiwavelength_2024, pestourie_inverse_2018, chen_gan_2017}. 


There are several promising avenues to further enhance MetaChat's capabilities. While our benchmark results demonstrate that prompt engineering is an effective strategy for AIM Agent development, fine-tuning LLMs with domain-specific literature  \cite{chung_scaling_2022, ouyang_training_2022, touvron_llama_2023} can significantly improve the capabilities and accuracy of AIM for complex, multi-step tasks. This is especially valuable in MetaChat-specific generalized reinforced policy optimization (GRPO) fine-tuning on long-term structured reasoning chains \cite{shao_deepseekmath_2024, deepseek-ai_deepseek-r1_2025} with tool use in the loop. Increased agentic design autonomy can allow the agent to design optimal figures of merit for achieving each unique user-queried design task. From a design interface perspective, multimodal LLMs can be integrated into the MetaChat framework to allow image-based instruction of the model to specify design preferences and constraints \cite{yin_survey_2024, openai_gpt-4_2024, ghafarollahi_atomagents_2024}.  We ultimately envision expansion of our multi-agentic and modular MetaChat framework to an ecosystem of specialized agents that supports  broader multi-scalar design capabilities within optics and photonics. To name a few examples, such agents can specialize in and utilize differential models for ray tracing, global optimization, and fabrication constraints.  We anticipate that through open-source initiatives in the photonics community \cite{jiang_metanet_2020, schubert_invrs-gym_2024}, MetaChat's capabilities can be continuously evaluated, expanded, and tuned to reflect the evolving design needs of practitioners in industry and academia \cite{chiang_chatbot_2024, zheng_lmsys-chat-1m_2024, srivastava_beyond_2023}.

The concept of augmenting specialized, LLM-driven agents with ultrafast and physically accurate surrogate solvers presents substantial opportunities for accelerating progress in applied physics research in both industry and academia.  For metasurfaces and nanophotonics in particular, MetaChat will particularly benefit adjacent research communities, ranging from astrophotonics atomic physics to AR/VR, where there is broad interest but limited expertise in implementing detailed nanophotonic solutions.
We anticipate even broader implications in the development and merging of ultrafast surrogate solvers with multi-agentic frameworks, where LLM agents can be used not only to generate novel scientific hypotheses \cite{gottweis_towards_2025}, but also to modify these hypotheses in real-time. These concepts extend beyond photonics and can be leveraged in multiphysics applications where agents can dynamically interleave solvers for different applied science domains to design entire novel systems. Furthermore, surrogate solver-augmented agentic frameworks can be coupled with robotic experimental environments, enabling automated end-to-end optimization of design and experimental validation \cite{roch_chemos_2018, dai_autonomous_2024, hatakeyama-sato_semiautomated_2024, woo_autonomous_2024}. This holds particularly exciting implications for the physical sciences as a whole, as it presents an unprecedented opportunity to push the boundaries of science using an infinite supply of ultrafast agentic intelligence.


\clearpage 

%

%
%
%
%
%
%


\section*{Acknowledgments}

We thank Dr. Mingkun Chen and Prof. You Zhou for technical assistance and helpful discussions.

\paragraph*{Funding:}
The work was supported by the National Aeronautics and Space Administration under Award 80NSSC21K0220 and the National Science Foundation under Award 2103301. R.~L. is supported by a graduate fellowship award from Knight-Hennessy Scholars at Stanford University.

\paragraph*{Author contributions:}
R.~L. and J.~A.~F. conceptualized the study. R.~L. coded and performed the analysis for MetaChat, AIM, FiLM WaveY-Net, and all related benchmarking. Y.~S. designed and coded the final optimization APIs. T.~D. and Y.~S. coded and implemented the differentiable neuroparameterization framework. R.~L., T.~D., Y.~S., and C.~M. contributed to the evaluation benchmark. K.~E. provided the mathematical formulation and initial code for the Stratton-Chu near-to-far field transformation. R.~L. coded the GPU-accelerated version of Stratton-Chu. R.~L. and J.~A.~F wrote the manuscript with input from all authors. J.~A.~F. supervised the research.

\paragraph*{Competing interests:}
There are no competing interests to declare.

\paragraph*{Data and materials availability:}
After peer review, all code will be open sourced and available at \url{https://github.com/jonfanlab/metachat}. The Stanford nanophotonics benchmark and the FiLM WaveY-Net training and evaluation datasets will be available in the Supplementary Information and at \url{http://metanet.stanford.edu/search/metachat} \cite{jiang_metanet_2020}, respectively. The Methods and Supplementary Information will be published after peer review.

\end{document}